\documentclass[twocolumn,showpacs,superscriptaddress,prb,amsmath,amssymb,letterpaper]{revtex4}
\usepackage{times}
\usepackage{amsmath,bm,amsfonts}
\usepackage{dcolumn}
\usepackage{graphicx}
\usepackage{latexsym}

\begin{document}

\title{Topological insulators and metal-insulator transition in the pyrochlore iridates}

\author{Bohm-Jung \surname{Yang}}
\affiliation{Department of Physics, University of Toronto,
Toronto, Ontario M5S 1A7, Canada}

\author{Yong Baek \surname{Kim}}
\affiliation{Department of Physics, University of Toronto,
Toronto, Ontario M5S 1A7, Canada}
\affiliation{School of Physics,
Korea Institute for Advanced Study, Seoul 130-722, Korea}

\date{\today}

\begin{abstract}
The possible existence of topological insulators in cubic pyrochlore iridates
A$_{2}$Ir$_{2}$O$_{7}$ (A = Y or rare-earth elements) is investigated by taking into account the strong
spin-orbit coupling and trigonal crystal field effect.
It is found that the trigonal crystal field effect, which is
always present in real systems,
may destabilize the topological insulator proposed for the ideal cubic crystal
field, leading to a metallic ground state. Thus the trigonal crystal field is an important
control parameter for the metal-insulator changeover.
We propose that this could be one of the reasons why distinct low temperature
ground states may arise for the pyrochlore iridates with different A-site ions.
On the other hand, examining the electron-lattice coupling, we find that
softening of the $\textbf{q}$=0 modes corresponding to trigonal
or tetragonal distortions of the Ir pyrochlore lattice
leads to the resurrection of the strong topological insulator.
Thus, in principle, a finite temperature transition to a low-temperature
topological insulator can occur via structural changes.
We also suggest that the application of the external pressure along [111]
or its equivalent directions would be the most efficient way of generating
strong topological insulators in pyrochlore iridates.
\end{abstract}

%\pacs{74.20.Mn, 74.25.Dw}

\maketitle

\section{\label{sec:intro} Introduction}

Topological band insulators arise from nontrivial Berry phase of electron
wave functions and possess gapless boundary states as a consequence of
topological properties of the bulk electron energy bands.\cite{Zhang_review,Kane_review,KaneMele1, KaneMele2,Roy1,Roy2,Fu-Kane-Mele,Moore,Fu-Kane,Fu-Kane2,bernevig1,bernevig2}
Topological invariants of electron wave functions in such topological insulators can be used to describe/identify
two dimensional quantum spin Hall insulators and three dimensional strong topological insulators.
The presence of the gapless boundary states and the associated topological
properties have recently been confirmed by a series of remarkable experiments
on HgCdTe, Bi$_{1-x}$Sb$_{x}$, Bi$_{2}$Se$_{3}$, Bi$_{2}$Te$_{3}$, and other materials.\cite{Exp0,Exp1,Exp2,Exp3,Exp4,Exp5}
It has been known that the strong spin-orbit coupling in these systems provides an essential
ingredient for the nontrivial Berry phase of electron wave functions.
One important future direction is to understand
the effect of interactions on the topological insulators.
Growing interests on combined effects of electron
correlation and topological properties have naturally lead to recent
fascinating studies on the realization of topological phases
in transition metal oxides with $d$ electrons,\cite{shitade,Pesin-Balents} instead of more conventional
$s$ or $p$ orbital systems where electron correlation is less important.

Transition metal oxides with 5$d$ electrons are characterized by the strong
spin-orbit coupling due to the large atomic number of 5$d$ transition metal elements.
As a result, the spin-orbit coupling competes with the kinetic and interaction energies, leading to
substantial correlation effects despite the relatively extended nature of
the 5$d$ orbitals. This feature, for example, is confirmed by recent experiments
on Sr$_{2}$IrO$_{4}$,\cite{bjkim1,bjkim2,hosub1,sjmoon1} where
the spin-orbit coupling plays an essential role in the formation of
the Mott-insulator ground state.
In the case of
the so-called hyperkagome material
Na$_4$Ir$_3$O$_8$, a spin liquid ground state is proposed,
manifesting strong correlation effects
in 5$d$ electron systems.~\cite{hyperkagome1,hyperkagome2,hyperkagome3,hyperkagome4,hyperkagome5,hyperkagome6,hyperkagome7,hyperkagome8}
Moreover iridium oxides are considered as promising candidate systems
where we can study the interplay between electron correlation and
strong spin-orbit coupling.
Perhaps most interestingly, the iridium oxides may also be
ideal materials for the occurrence of topological insulators.
It is suggested that
the honeycomb lattice of Ir ions in Na$_{2}$IrO$_{3}$ with the complex
hopping amplitudes arising from the strong spin-orbit coupling may lead
to a quantum spin Hall insulator.\cite{shitade}
Three dimensional pyrochlore lattice of Ir
is also suggested as a candidate system for a strong topological insulator.~\cite{Pesin-Balents}
The possibility of the topological Mott insulators where the spinons,
not the electrons, possess topological band structures is discussed as a result
of correlation effects.~\cite{sslee,Pesin-Balents}

%%%%%%%%%%%%%%%%%%%%%%%%%%%%%%%%%%%%%
\begin{figure}[t]
\centering
\includegraphics[width=8.5 cm]{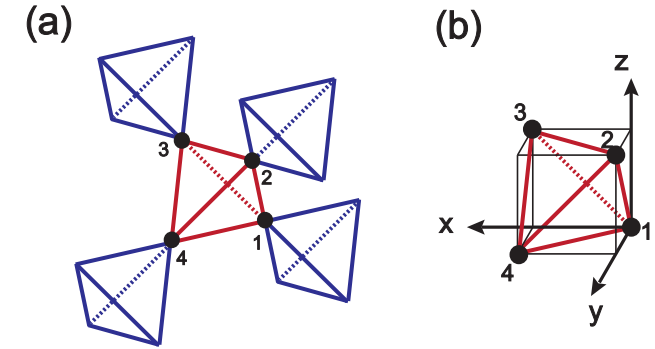}
\caption{
(a) Structure of the pyrochlore lattice.
An Ir atom sits at each vertex of a tetrahedron.
Four sites within a tetrahedral unit cell are numbered from 1 to 4.
(b) The coordinate system used in this paper.
}
\label{fig:pyrochlorestructure}
\end{figure}
%%%%%%%%%%%%%%%%%%%%%%%%%%%%%%%%%%%%%%%%%%%

In this work, we study the possibility and stability of strong topological insulators in
pyrochlore iridates, A$_{2}$Ir$_{2}$O$_{7}$ (A = Y or rare-earth elements) by taking into
account both of the strong spin-orbit coupling and trigonal crystal field effect induced
by a local distortion of IrO$_{6}$ octahedra.
Due to the extended nature of 5$d$ electron wave functions,
the local crystal field strength in 5$d$ systems also becomes a large energy
scale competing with electron interaction and spin-orbit coupling.~\cite{hosub2}
The strength of the trigonal crystal field effect may be different for distinct
choices of the A-site ion in A$_{2}$Ir$_{2}$O$_{7}$. Given that there exist a variety
of choices for the A-site, the pyrochlore iridates offer the opportunity to control
the relative strength of the crystal field effect and the spin-orbit coupling.
In addition, the pyrochlore iridates show both metallic and insulating ground
states at low temperatures as well as finite temperature metal-insulator transitions
depending on the A-site ions.~\cite{gingras,MIT,MIT0} Thus it would be very interesting to understand
the origin of distinct phases and phase transitions.

Here we derive an effective Hamiltonian for the Ir ions in the pyrochlore iridates
by carefully incorporating the trigonal crystal field effect and using the appropriate
spin-orbital basis. Previously an effective Hamiltonian was derived by Pesin and
Balents~\cite{Pesin-Balents} for the Ir ions in a perfect octahedral environment with the cubic $O_{h}$ symmetry
and it was found that the Ir pyrochlore system supports a strong topological insulator
when the strength of spin-orbit coupling is sufficiently large.
In contrast we find that the trigonal crystal field effect, which is always present in
real pyrochlore systems, is quite significant and it can lead to a metallic ground state.

On the other hand, we find that the distortions of the Ir pyrochlore lattice
induced by softening of certain $\textbf{q}$=0 phonon modes give rise to the resurrection of
strong topological insulators. This suggests that finite temperature metal-insuator
transition can occur through structural distortions such that the low temperature
ground state is a strong topological insulator. It is interesting to notice that
recent experiments on Sm$_{2}$Ir$_{2}$O$_{7}$ and Eu$_{2}$Ir$_{2}$O$_{7}$
show structural changes at the metal-insulator transition even though
the low temperature ground state seems to be magnetic and show spin-glass-like
behavior.~\cite{Raman,Magnetic}
We also suggest from the studies of the electron-lattice coupling that
applying pressure along [111]
or its equivalent directions would be the most efficient way of generating
strong topological insulators in pyrochlore iridates.

The rest of the paper is organized as follows.
In Sec.~\ref{sec:effectiveHamiltonian} we derive
the effective hopping Hamiltonian including the trigonal
crystal field effect. We first consider the influence
of local trigonal distortion of the oxygen octahedra on Ir $t_{2g}$ electrons.
Then we explain the derivation of the corresponding lattice Hamiltonian.
In Sec.~\ref{sec:trigonal_band},
the evolution of the electronic structure induced by trigonal crystal field in
the presence of the strong spin-orbit coupling is described in detail.
Here we explain how a metallic state may arise when the trigonal crystal field
effect is present.
In Sec.~\ref{sec:LatticeDistortion}, we discuss the effect of electron-lattice
coupling on the electronic structure of the Ir pyrochlore system.
We show how the electron-lattice coupling can lead to strong topological
insulators.
Finally, we conclude in Sec.\ref{sec:conclusion}.
The detailed expressions for various matrices describing
the effective Hamiltonian are given in the Appendix.

\section{\label{sec:effectiveHamiltonian} Effective hopping Hamiltonian with trigonal crystal field}
In this section we investigate the effect of local trigonal crystal field
on the electronic structure of Ir 5$d$ electrons.
In Sec.~\ref{sec:trigonal} we explain the degeneracy lifting of $t_{2g}$ orbitals
under trigonal crystal fields.
The effective hopping Hamiltonian including the trigonal crystal field
is derived in Sec.~\ref{sec:H_hopping}.
Detailed procedures of the derivation can be found there, as well.

\subsection{\label{sec:trigonal} Trigonal crystal field}

\begin{figure}[t]
\centering
\includegraphics[width=6.5 cm]{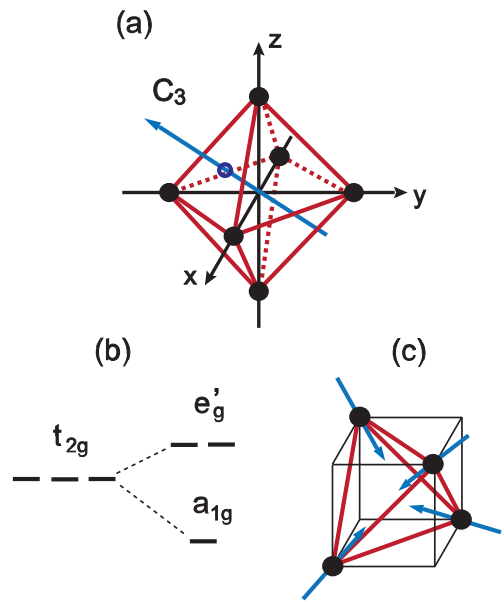}
\caption{
Trigonal crystal field around an $\text{Ir}^{4+}$ ion.
(a) Local geometry of an IrO$_{6}$ octahedron.
A trigonal distortion is induced by compression or elongation
of the surrounding oxygen octahedron along
a $C_{3}$ symmetry axis for 3-fold rotation.
(b) Splitting of degenerate $t_{2g}$ states
under the trigonal crystal field.
(c) Local $C_{3}$ axes for four Ir ions constituting a
tetrahedron; Each $C_{3}$ axis points to the body center
of the tetrahedron.
}
\label{fig:trigonalcrystalfield}
\end{figure}

Each $\text{Ir}^{4+}$ ion is coordinated by six oxygen anions which
are at equal distance from the central $\text{Ir}^{4+}$ cation.
The actual coordinates of the surrounding oxygen anions have a free positional parameter,
so-called the oxygen $x$ parameter, which depends on material properties.
In general, the six oxygen ions around a central $\text{Ir}^{4+}$ form a distorted octahedron
where the amount of distortion can be parameterized
by the oxygen $x$ parameter.\cite{Structure_Hiroi, Structure_Review} For $x=x_c$ = 5/16,
each $\text{Ir}^{4+}$ ion is under a perfect local cubic crystal field.
The deviation of $x$ from the ideal value of $x_c$
generates a trigonal crystal field.
The trigonal distortion of an oxygen octahedron is accompanied
by compression ($x>x_{c}$) or elongation ($x<x_{c}$) of the oxygen octahedron
along one of the $C_{3}$ symmetry axis for 3-fold rotation.
A local geometry of an IrO$_{6}$ octahedron and a $C_{3}$ symmetry axis
for the trigonal distortion are described in Fig.~\ref{fig:trigonalcrystalfield}(a).

%%%%%%%%%%%%%%%%%%%%%%%%%%%%%%%%%%%%%
%%%%%%%%%%%%%%%%%%%%%%%%%%%%%%%%%%%%%%%%%%%

Under the trigonal distortion,
the local symmetry group around an Ir site reduces from the cubic $O_{h}$
to $D_{3d}$.
Due to the trigonal crystal field, degenerate $t_{2g}$ orbitals split into
a singlet $a_{1g}$ state and a doublet $e'_{g}$ state
belonging to $A_{1g}$ and $E_{g}$
irreducible representations of the $D_{3d}$ point group, respectively.
(See Fig.~\ref{fig:trigonalcrystalfield}(b).)
A $C_{3}$ symmetry axis corresponds to one of the [111] or its equivalent axes.
Taking the local $[n_{1}n_{2}n_{3}]$ directions
as the $C_{3}$ symmetry axes ($n_{i}=\pm1$),
the effect of trigonal crystal field on $t_{2g}$ states
can be described by
the following Hamiltonian,

\begin{displaymath}
H_{\text{tri}}=-\frac{\Delta_{\text{tri}}}{3}
\left( \begin{array}{ccc}
0 & n_{1}n_{2} & n_{1}n_{3}  \\
n_{1}n_{2} & 0 &  n_{2}n_{3}  \\
n_{1}n_{3} &  n_{2}n_{3} & 0
\end{array} \right).
\end{displaymath}
Here we have chosen a local basis as $\Psi^{\dag}=(d^{\dag}_{yz},d^{\dag}_{zx},d^{\dag}_{xy})$.
The singlet $a_{1g}$ state with the energy $E_{a}=-2\Delta_{\text{tri}}$/3 is given by
\begin{align}\label{eq:trigonal1}
|a_{1}\rangle \equiv a^{\dag}_{1}|0\rangle
=\frac{1}{\sqrt{3}}[n_{1}|d_{yz}\rangle + n_{2}|d_{zx} \rangle + n_{3}|d_{xy}\rangle].
\end{align}
The eigenvectors for the $e'_{g}$ doublet states with the energy $E_{e}=\Delta_{\text{tri}}$/3
can be written as
\begin{align}\label{eq:trigonal2}
|e_{+}\rangle &\equiv e^{\dag}_{+}|0\rangle
=\frac{1}{\sqrt{3}}[n_{1}\omega|d_{yz}\rangle + n_{2}\omega^{2}|d_{zx} \rangle + n_{3}|d_{xy}\rangle],
\nonumber\\
|e_{-}\rangle &\equiv e^{\dag}_{-}|0\rangle
=\frac{1}{\sqrt{3}}[n_{1}\omega^{2}|d_{yz}\rangle + n_{2}\omega|d_{zx} \rangle + n_{3}|d_{xy}\rangle],
\end{align}
where $\omega=e^{i\frac{2\pi}{3}}$.

In Fig.~\ref{fig:trigonalcrystalfield}(c),
we show the local $C_{3}$ axes of the four Ir sites
constituting a tetrahedral unit cell.
Each $C_{3}$ axis points to the body center of the tetrahedron.
Considering the trigonal distortions of IrO$_{6}$ octahedra and their
relative orientations, we construct the effective hopping
Hamiltonian for Ir $d$ electrons on the pyrochlore lattice.
The resulting Hamiltonian is given by
\begin{align}\label{eq:finalHamiltonian}
H=&\sum_{i,n,\alpha}(\varepsilon_{\alpha}-\mu)d^{\dag}_{in\alpha}d_{in\alpha}
\\\nonumber
&+\sum_{\langle ij \rangle}\sum_{n,n'}\sum_{\alpha,\alpha'}
\{d^{\dag}_{in\alpha}T_{n\alpha,n'\alpha'}d_{jn'\alpha'} + h.c.\}.
\end{align}
Here $i$ and $j$ are indices for unit cells while $n$ and $n'$ indicate
Ir sites within a unit cell ($n=1,2,3,4$).
We use the index $\alpha$ ($\alpha=1,...6$) to describe local spin-orbit eigenstates.
Within the $t_{2g}$ manifold, $d$ electrons behave
as if their effective orbital angular
momentum is one with an additional minus sign.\cite{Pesin-Balents}
Therefore the effective total angular momentum of Ir $d$ electrons, $j_{\text{eff}}$
can be either 1/2 or 3/2.
We use $\alpha=1,2$ to indicate the spin-orbit doublet with $j_{\text{eff}}=1/2$
and $\alpha=3,4,5,6$ to denote the spin-orbit quadruplet with $j_{\text{eff}}=3/2$.
The on-site energy $\varepsilon_{\alpha}=\lambda_{SO}$ for the spin-orbit doublet and
$\varepsilon_{\alpha}=-\lambda_{SO}/2$ for the spin-orbit quadruplet with
the spin-orbit coupling strength given by $\lambda_{SO}$.
$T_{n\alpha,n'\alpha'}$ is the hopping amplitudes between nearest neighbor Ir sites
and $\mu$ is the chemical potential.

In contrast to the tight-binding Hamiltonian
derived by Pesin and Balents in Ref.\onlinecite{Pesin-Balents}
where a single energy scale $t$ describes the hopping processes,
the hopping amplitude $T_{n\alpha,n'\alpha'}$  has two independent parameters $t_{a}$ and $t_{e}$
in this case.
These two hopping parameters come from the hopping processes
between $a_{1g}$ and $e'_{2g}$ states under the trigonal crystal field, respectively.
Under the perfect local cubic environment around an Ir site without a trigonal crystal field,
$t_{a}=t_{e}$ is satisfied.
In other words, the effect of trigonal crystal fields on the electronic structure
can be captured by varying the relative magnitude of $t_{a}/t_{e}$.
In the following section, we explain the procedures deriving the above effective
Hamiltonian in Eq.(\ref{eq:finalHamiltonian}).

\subsection{\label{sec:H_hopping} Construction of the effective hopping Hamiltonian}

%%%%%%%%%%%%%%%%%%%%%%%%%%%%%%%%%%%%%
\begin{figure}[t]
\centering
\includegraphics[width=8 cm]{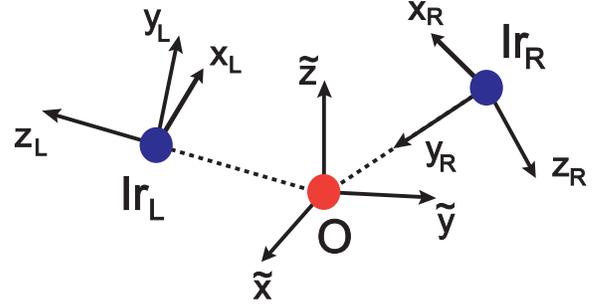}
\caption{
Local Ir-O-Ir configuration.
Local octahedral axes for Ir atoms on the left-hand side (Ir$_{L}$)
and the right-hand side (Ir$_{R}$) are not parallel to
the global cubic axis adopted for the central oxygen atom.
}
\label{fig:threeatom}
\end{figure}
%%%%%%%%%%%%%%%%%%%%%%%%%%%%%%%%%%%%%%%%%%%

We construct the effective Hamiltonian describing hopping processes
between neighboring Ir $t_{2g}$ orbitals
through intermediating oxygen $p$ orbitals in the following way.
We first consider a small cluster composed of neighboring Ir-O-Ir atoms
as shown in Fig.~\ref{fig:threeatom}.
Allowing Ir-O hopping processes, the three-atom cluster can be described
by the following Hamiltonian,

\begin{align}
H = H^{(0)} + H^{(1)},
\end{align}
where
\begin{align}
H^{(0)} = \eta^{\dag}_{L}H^{(0)}_{d}\eta_{L} + \eta^{\dag}_{R}H^{(0)}_{d}\eta_{R} + \tilde{p}^{\dag}H^{(0)}_{p}\tilde{p},
\end{align}
and
\begin{align}
H^{(1)} = \eta^{\dag}_{L}V_{L}\tilde{p} + \tilde{p}^{\dag}V^{\dag}_{L}\eta_{L}
+ \eta^{\dag}_{R}V_{R}\tilde{p} + \tilde{p}^{\dag}V^{\dag}_{R}\eta_{R}.
\end{align}

Here $H^{(0)}$ indicates on-site potentials at Ir and O sites
while $H^{(1)}$ describes Ir-O hybridizations.
The on-site atomic potentials of $d$ and $p$ orbitals give rise to
the diagonal matrices, $H^{(0)}_{d}=\text{diag}[E_{a}, E_{e}, E_{e}]$
and  $H^{(0)}_{p}=\text{diag}[E_{p}, E_{p}, E_{p}]$.
Here $E_{a}$ ($E_{e}$) is the atomic energy for an $a_{1g}$ singlet ($e'_{g}$ doublet) state
composed of Ir $d$ orbitals under the trigonal crystal field
and $E_{p}$ is the atomic energy of degenerate oxygen $p$ orbitals.
$V_{n}$ ($n$= $L$, $R$) represents overlap integrals between Ir
$d$ orbitals and oxygen $p$ orbitals.
Taking into account the local trigonal crystal field of each Ir atom,
we employ a basis, that is diagonal under the local trigonal crystal field,
$\eta^{\dag}_{n}=(a^{\dag}_{1,n}, e^{\dag}_{+,n}, e^{\dag}_{-,n})$
to represent Ir $t_{2g}$ orbitals.
The subscript $n$ indicates
the locations of two Ir atoms
on the left-hand side ($n=L$) or the right-hand side ($n=R$).
$\tilde{p}^{\dag}=(\tilde{p}^{\dag}_{x}, \tilde{p}^{\dag}_{y}, \tilde{p}^{\dag}_{z})$ represents
oxygen $p$ orbitals connecting neighboring Ir atoms.
It is important to notice that the local octahedral axes of Ir atoms
are not parallel to the global cubic axes as described in Fig.~\ref{fig:threeatom}.
Each component of the oxygen $p$ orbitals is defined with respect to
the global cubic axis, while $d$ orbital basis is defined with
respect to the local octahedral coordinate of each Ir atom.

The effective hopping Hamiltonian between two neighboring Ir atoms
is obtained from the second order perturbation theory
treating $H^{(1)}$ as a perturbation. The resulting
Hamiltonian is given by
\begin{align}\label{eq:Heff1}
&H_{L,R} =
\nonumber\\
&-\frac{1}{2}\eta^{\dag}_{L}\Big\{\frac{1}{H^{(0)}_{p}-H^{(0)}_{d}}V_{L}V^{\dag}_{R}
+V_{L}V^{\dag}_{R}\frac{1}{H^{(0)}_{p}-H^{(0)}_{d}}\Big\}\eta_{R} + h.c.,
\end{align}

The information about the relative orientation between two local octahedral axes of neighboring
Ir atoms is contained in $V_{L}$ and $V_{R}$.
To construct $V_{L}$ and $V_{R}$, we follow the approach
used by Pesin and Balents in Ref.~\onlinecite{Pesin-Balents}.
We first construct matrices describing hopping processes
between Ir $t_{2g}$ and O $p$ orbitals.
It is convenient to define components of
oxygen $p$ and Ir $t_{2g}$ orbitals with respect to the same cubic axis.
In this case the Ir-O hopping processes are
strongly constrained by the symmetries of $d$ and $p$ orbitals.
The hopping matrices connecting ($p_{x}$, $p_{y}$, $p_{z}$) orbitals to
($d_{yz}$, $d_{zx}$, $d_{xy}$) orbitals are given by~\cite{Pesin-Balents}

\begin{displaymath}
\tau^{\pm}_{x} =
\pm \alpha_{pd\pi}
\left( \begin{array}{ccc}
0 & 0 & 0  \\
0 & 0 & 1  \\
0 & 1 & 0
\end{array} \right),
\end{displaymath}

\begin{displaymath}
\tau^{\pm}_{y} =
\pm \alpha_{pd\pi}
\left( \begin{array}{ccc}
0 & 0 & 1  \\
0 & 0 & 0  \\
1 & 0 & 0
\end{array} \right),
\end{displaymath}

\begin{displaymath}
\tau^{\pm}_{z} =
\pm \alpha_{pd\pi}
\left( \begin{array}{ccc}
0 & 1 & 0  \\
1 & 0 & 0  \\
0 & 0 & 0
\end{array} \right).
\end{displaymath}

Here the subscript and superscript of the matrix $\tau$
indicate the location of an O atom with respect to the Ir atom.
For example, $\tau^{+}_{x}$ describes Ir-O hopping processes when an O atom
sits on the positive x axis with the Ir atom sitting at the origin.
$\alpha_{pd\pi}$ is the hopping integral between neighboring $p$ and $d$ orbitals.

Since the local octahedral axes of Ir atoms
are not parallel, we introduce an additional matrix $R^{(i)}$
for each $i$th Ir atom,
which rotates the global cubic axis to a local octahedral axis.
Under the action of $R^{(i)}$, the $\tilde{p}_{n}$ ($n$ = $x,y,z$) orbitals
defined with respect to the global cubic axis
transform as $p^{\dag}_{i,n}=R^{(i)}_{mn}\tilde{p}^{\dag}_{m}$.
Here $p_{i,n}$ denotes a $p$ orbital defined with respect to
the local octahedral axis of the $i$th Ir atom.
In addition, we introduce the matrix $\Omega$ that transforms
($d_{yz}$, $d_{zx}$, $d_{xy}$) to the trigonal basis ($a_{1}$,
$e_{+}$, $e_{-}$) in the following way, $e^{\dag}_{i,n}=\Omega_{nm}d^{\dag}_{i,m}$.
From Eq.~(\ref{eq:trigonal1}) and Eq.~(\ref{eq:trigonal2}),
$\Omega$ is obtained as,

\begin{displaymath}
\Omega =
\frac{1}{\sqrt{3}}
\left( \begin{array}{ccc}
n_{1} & n_{2} & n_{3}  \\
n_{1} \omega & n_{2} \omega^{2} & n_{3}  \\
n_{1} \omega^{2} & n_{2} \omega & n_{3}
\end{array} \right).
\end{displaymath}
Now we can construct $V_{L}$ and $V_{R}$ explicitly, which are given by
$V_{L}=\Omega^{*}\tau_{p(L)}R^{(L)\dag}$ and
$V_{R}=\Omega^{*}\tau_{p(R)}R^{(R)\dag}$.
Here the subscript $p(L)$ ($p(R)$) of $\tau$ indicates the location of
the intermediating oxygen atom with respect to
the local octahedral axis of the Ir ion on the left-hand (right-hand) side.

The effective hopping Hamiltonian in Eq.(\ref{eq:Heff1})
can be written compactly as,

\begin{align}
&H_{L,R} =
\frac{1}{2}\eta^{\dag}_{L}\Big\{T_{\text{hop}}V_{L}V^{\dag}_{R}
+V_{L}V^{\dag}_{R}T_{\text{hop}}\Big\}\eta_{R} + h.c.,
\end{align}
where $T_{\text{hop}}$$\equiv$diag$[t_{a}, t_{e}, t_{e}]$
with $t_{a}=\alpha^{2}_{pd\pi}/(E_{a}-E_{p})$ and
$t_{e}=\alpha^{2}_{pd\pi}/(E_{e}-E_{p})$. In the above,
we have replaced $V_{R}$ ($V_{L}$) by $V_{R}/\alpha_{pd\pi}$
($V_{L}/\alpha_{pd\pi}$) to make $t_{a}$ and $t_{e}$ have
conventional forms for hopping amplitudes.
Note that $t_{a}=t_{e}$ in the absence of the trigonal crystal field effects
because $E_{a}=E_{e}$ is satisfied in that case.
The effect of local trigonal field splitting can be considered
by changing the relative magnitude of $t_{a}/t_{e}$.

Up to now we have neglected spin degrees of freedom.
Since the spin-orbit coupling is the largest energy scale of
the problem, it is convenient to use the local spin-orbit eigenstates
as a basis for the representation of the Hamiltonian.
To project the Hamiltonian onto the local spin-orbit basis, we define rotation matrices $D^{(i)}$
acting on the spin space, which is nothing but a spinor representation
of $O(3)$ rotations $R^{(i)}$ defined before.
The detailed expressions of matrices $R^{(i)}$ and $D^{(i)}$
are given in the Appendix.
Then the spin dependent hopping Hamiltonian is given by

\begin{align}
&H_{L,R} =\sum_{l\sigma,l'\sigma'}\eta^{\dag}_{L,l\sigma}\tilde{T}_{l\sigma,l'\sigma'}\eta_{R,l'\sigma'} + h.c.,
\end{align}
where

\begin{align}\label{eq:Ttilde}
&\tilde{T}_{l\sigma,l'\sigma'}=\frac{1}{2}[T_{\text{hop}}V_{L}V^{\dag}_{R}
+V_{L}V^{\dag}_{R}T_{\text{hop}}]_{l,l'}[(D^{(L)})^{\dag}D^{(R)}]_{\sigma,\sigma'},
\end{align}
Here $l$ is the index for $d$-orbitals under the local trigonal crystal field
($l$ = $a$, $e_{+}$, $e_{-}$)
and $\sigma$ indicates the spin projection with respect to a local quantization axis.

We can obtain the final expression for the effective hopping Hamiltonian
by introducing a matrix $A$ which changes the trigonal basis to
the local spin-orbit basis in the following way,
$|\alpha\rangle = d^{\dag}_{\alpha}|0\rangle\equiv\sum_{l\sigma}A_{\alpha,l\sigma}\eta^{\dag}_{l\sigma}|0\rangle$.
Here we denote the local spin-orbit basis using an index $\alpha$ ($\alpha=1,...6$).
$\alpha=1,2$ indicate the spin-orbit doublet with the total angular momentum $j_{\text{eff}}=1/2$
and $\alpha=3,4,5,6$ denote the spin-orbit quadruplet with $j_{\text{eff}}=3/2$.
It is straight forward to extend the approach described above for a three-site cluster
to the full pyrochlore lattice system.
The resulting effective lattice hopping Hamiltonian is given by
\begin{align}
&H_{\text{eff}} =\sum_{\langle ij \rangle}\sum_{n,n'}\sum_{\alpha,\alpha'}
d^{\dag}_{in\alpha}T_{n\alpha,n'\alpha'}d_{jn'\alpha'} + h.c.,
\end{align}
where

\begin{align}\label{eq:Tfinal}
&T_{n\alpha,n'\alpha'}=\sum_{l\sigma,l'\sigma'}(A^{*})_{\alpha,l\sigma}\tilde{T}_{nl\sigma,n'l'\sigma'}(A^{T})_{l'\sigma',\alpha'}.
\end{align}
Here $i$ is the unit cell index and the index $n$ refers to the four Ir sites
within a single tetrahedral unit cell.
Including the on-site potentials for the local spin-orbit eigenstates,
the expression for the effective tight-binding Hamiltonian is finally given by
Eq.(\ref{eq:finalHamiltonian}).

\section{\label{sec:trigonal_band} Evolution of the electronic structure under trigonal crystal field }

%%%%%%%%%%%%%%%%%%%%%%%%%%%%%%%%%%%%%
\begin{figure}[t]
\centering
\includegraphics[width=8 cm]{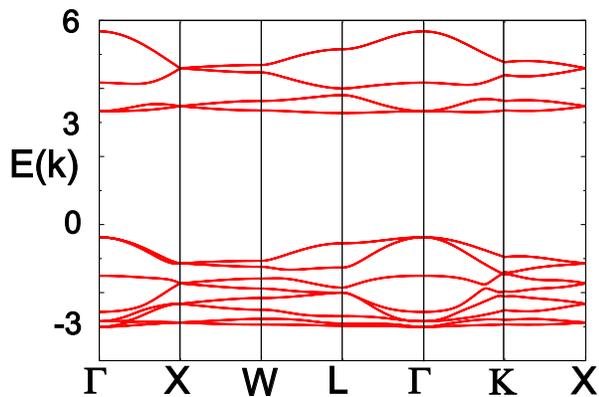}
\caption{
Tight-binding bandstructure of Ir 5$d$ orbitals
when there is no trigonal crystal field effect.
Here we have chosen $\lambda_{SO}$=4.0, $t_{a}$=$t_{e}$=0.5.
Note that the upper 4 bands corresponding to the spin-orbit
doublet ($j_{\text{eff}}$=1/2) are well separated from the other 8 bands which come
from the spin-orbit quadruplet ($j_{\text{eff}}$=3/2).
}
\label{fig:fullband}
\end{figure}
%%%%%%%%%%%%%%%%%%%%%%%%%%%%%%%%%%%%%%%%%%%

In this section, we discuss the evolution of the electronic structure of the Ir 5d system on
the pyrochlore lattice in the presence of the trigonal crystal field.
We start with the case where the Ir ions are under the perfect cubic crystal field with
large spin-orbit coupling.
In Fig.~\ref{fig:fullband}, we plot the electron band dispersion along the high symmetry directions
in the Brillouin zone for the perfect cubic crystal field on the Ir ions with
$\lambda_{SO}$=4.0, $t_{a}$=$t_{e}$=0.5.
Since we have 4 sites within a unit cell and each site
supports 3 $t_{2g}$ orbitals, there are 24 bands within
the first Brioullin zone including spin degrees of freedom.
Due to the time reversal and inversion symmetries,
each band is doubly degenerate. In Fig.~\ref{fig:fullband},
the upper 4 bands
are derived from the spin-orbit doublets with
total angular momentum $j_{\text{eff}}$=1/2.
On the other hand, the lower 8 bands
come from the quadruplets with $j_{\text{eff}}$=3/2.
Note that these two groups of bands are well separated by
a large energy gap with an energy scale given by $\lambda_{SO}$.

\begin{figure}[t]
\centering
\includegraphics[width=8.5 cm]{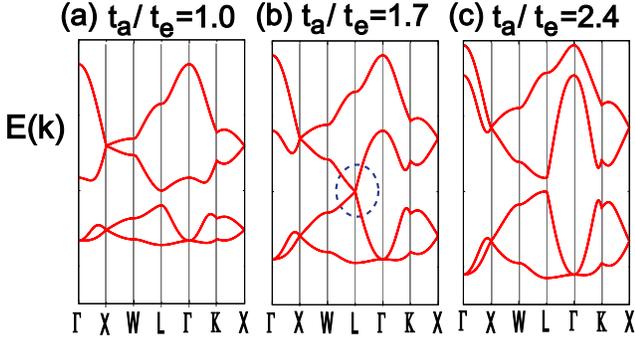}
\caption{
Dispersions of $j_{\text{eff}}$=1/2 bands under
trigonal crystal field with $t_{a} > t_{e}$.
Here we have fixed $\lambda_{SO}=4.0$, $t_{e}=0.5$ while changing $t_{a}$.
There is an accidental band touching between
the two bands in the middle at the $L$ point when $t_{a}/t_{e}=1.7$.
}
\label{fig:trigonalband1}
\end{figure}

Since each Ir atom contributes 5 electrons,
20 bands among the 24 bands are filled.
Namely, we have a band insulator with the half-filled $j_{\text{eff}}$=1/2 bands.
Therefore in the forthcoming discussion, we neglect
the fully occupied $j_{\text{eff}}$=3/2 bands and focus on
the properties of the upper 4 bands (or 8 bands counting the double degeneracy of each band)
possessing $j_{\text{eff}}$=1/2 character.
The energy dispersions of the $j_{\text{eff}}=1/2$ states are shown in Fig.~\ref{fig:trigonalband1}(a).
The fully occupied lower two bands are separated from the upper two bands by a finite gap between them.

To understand the topological properties of the insulating phase,
we compute the $Z_{2}$ topological invariants $(\nu;\nu_{1}\nu_{2}\nu_{3})$
from the parity eigenvalues $\xi_{m}(\bf{\Gamma}_{l})$ at the time reversal invariant momenta,
following Fu and Kane.~\cite{Fu-Kane}
Here $\xi_{m}(\bf{\Gamma}_{l})$ indicates the inversion parity of the $m$th occupied $j_{\text{eff}}=1/2$
band at the time-reversal invariant momentum $\bf{\Gamma}_{l}$.
Using the reciprocal lattice vectors $\textbf{G}_{i}$ ($i$=1, 2, 3),
the eight time reversal invariant momenta can be written as $\bf{\Gamma}_{l=n_{1}n_{2}n_{3}}$
=$(n_{1}\textbf{G}_{1}+n_{2}\textbf{G}_{2}+n_{3}\textbf{G}_{3})/2$ with $n_{1,2,3}= 0, 1$.
The strong $Z_{2}$ topological invariant $\nu$ is given by
\begin{align}\label{eq:strongnu}
(-1)^{\nu}=\prod_{n_{i}=0,1}\prod_{m=1}^{2}\xi_{m}(\bf{\Gamma}_{n_{1}n_{2}n_{3}}),
\end{align}
where the parity eigenvalues at the eight time reversal invariant momenta are multiplied
at the same time.
On the other hand, each of the three weak $Z_{2}$ topological
invariants $\nu_{i}$ ($i$=1,2,3) is determined by the parity eigenvalues at the four time reversal invariant momenta
lying on a plane,
which is given by
\begin{align}\label{eq:weaknu}
(-1)^{\nu_{i}}=\prod_{n_{i}=1, n_{j\neq i}=0,1}\prod_{m=1}^{2}\xi_{m}(\bf{\Gamma}_{n_{1}n_{2}n_{3}}).
\end{align}
Because of the time reversal symmetry, each band is doubly degenerate at the time reversal
invariant momentum and every Kramers doublet share the same inversion parity.
Since the $Z_{2}$ topological invariants count the parity of one state for each Kramers pair,~\cite{Fu-Kane}
we consider the product of the inversion parities corresponding to the two occupied $j_{\text{eff}}=1/2$ bands
in Eq.(\ref{eq:strongnu}) and (\ref{eq:weaknu}).
Notice that, since the product of the inversion parities of the occupied $j_{\text{eff}}=3/2$ bands is
+1 in every time-reversal-invariant momentum, we can neglect the contributions from the $j_{\text{eff}}=3/2$
bands.
These analyses lead to a strong topological insulator with the $Z_{2}$ invariants
(1;000) as found earlier by Pesin and Balents.~\cite{Pesin-Balents}
It is interesting to note that a strong topological insulator
with the same $Z_{2}$ invariant (1;000) was also found
in a simple one-band model on the pyrochlore lattice.~\cite{franz}

%%%%%%%%%%%%%%%%%%%%%%%%%%%%%%%%%%%%%
%%%%%%%%%%%%%%%%%%%%%%%%%%%%%%%%%%%%%%%%%%%

\begin{table}
\begin{tabular}{@{}|c|c|c|c|c|}
\hline \hline
& $E_{1}$ & $E_{2}$ & $E_{3}$ & $E_{4}$ \\
\hline \hline
$\Gamma(0,0,0)$ & + & + & + & +  \\
$L_{0}(\pi,\pi,\pi)$ & - & - & + & - \\
$L_{1}(-\pi,\pi,\pi)$ & + & + & - & + \\
$L_{2}(\pi,-\pi,\pi)$ & + & + & - & + \\
$L_{3}(\pi,\pi,-\pi)$ & + & + & - & + \\
$X_{1}(2\pi,0,0)$ & $\epsilon$ & $\bar{\epsilon}$ & $\epsilon'$ & $\bar{\epsilon'}$ \\
$X_{2}(0,2\pi,0)$ & $\epsilon$ & $\bar{\epsilon}$ & $\epsilon'$ & $\bar{\epsilon'}$ \\
$X_{3}(0,0,2\pi)$ & $\epsilon$ & $\bar{\epsilon}$ & $\epsilon'$ & $\bar{\epsilon'}$ \\
\hline \hline
\end{tabular}
\caption{Inversion parities of the $j_{\text{eff}}$=1/2 bands at time reversal invariant momenta
for $\lambda_{SO}=4.0$, $t_{a}=t_{e}=0.5$. Here $E_{1}\leq E_{2} \leq E_{3} \leq E_{4}$.
At the momentum $X_{i}$($i$=1, 2, 3), the upper (lower) two bands are degenerate
with opposite inversion parities satisfying $\epsilon\bar{\epsilon}=-1$ ($\epsilon'\bar{\epsilon'}=-1$). }
\label{table:bandparity}
\end{table}

Now we describe the effect of the trigonal
crystal field on the electronic structure of the $j_{\text{eff}}$=1/2 bands.
As mentioned above, the trigonal crystal field effect can be described by changing
the relative magnitude of $t_{a}$ and $t_{e}$.
In general, the pyrochlore oxides have oxygen $x$ parameters
ranging from 0.309 to 0.355.~\cite{Structure_Review}
Since $x_{c}$=0.3125
for the perfect cubic crystal field,
we have to consider both trigonal compression ($x>x_{c}$)
and elongation ($x<x_{c}$) cases.
According to the naive crystal field splitting picture,
the on-site energies $E_{a}$ and $E_{e}$ of $a_{1g}$ and $e'_{g}$ states
are determined by the magnitude of oxygen $x$ parameters.
However, in real materials, the relative magnitude between $E_{a}$
and $E_{e}$ are strongly affected by hybridization with
high energy $e_{g}$ orbital states, which are allowed
under the trigonal crystal field.~\cite{landron}
Therefore irrespective of the magnitude of the oxygen $x$ parameter,
we have to investigate both $t_{a}/t_{e}>1$ and $t_{e}/t_{a}>1$
cases on equal footing.

We first consider the case of $t_{a}/t_{e}>1$.
In Fig.~\ref{fig:trigonalband1}, we plot the evolution
of the band structure as we increase $t_{a}/t_{e}$.
For $t_{a}/t_{e}\approx 1.7$, an accidental band touching occurs
at the $L$ point in the Brillouin zone.
Since the two bands touching at the $L$ point have opposite
inversion parities as shown in Table~\ref{table:bandparity},
the band crossing induces the exchange
of the parities between the two band touching at the $L$ point.
However, since we have eight different momentum points  within the first Brillouin zone,
which are symmetry equivalent to the $L$ point,
the product of the inversion parities for all occupied bands is invariant.
According to Fu and Kane,~\cite{Fu-Kane} the product of the inversion parities of occupied bands
determines the $Z_{2}$ topological invariant $\nu$, characterizing the topological properties of
insulators.
Therefore the accidental band touching does not induce the change of the topological properties
of the insulating states.

%%%%%%%%%%%%%%%%%%%%%%%%%%%%%%%%%%%%%
\begin{figure}[t]
\centering
\includegraphics[width=8.5 cm]{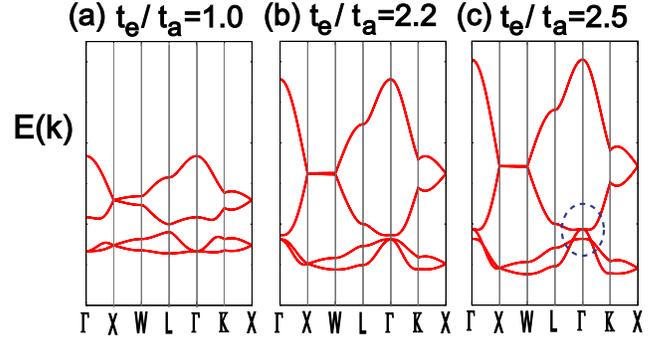}
\caption{
Dispersions of $j_{\text{eff}}$=1/2 bands under
trigonal crystal field with $t_{a} < t_{e}$.
Here we have fixed $\lambda_{SO}=4.0$, $t_{a}=0.5$ but vary $t_{e}$.
As $t_{e}/t_{a}$ increases the energy gap between
the two bands in the middle reduces. When $t_{e}/t_{a}\approx 2.3$
band inversion occurs at the $\Gamma$ point and
the system becomes metallic.
}
\label{fig:trigonalband2}
\end{figure}
%%%%%%%%%%%%%%%%%%%%%%%%%%%%%%%%%%%%%%%%%%%

Now we consider the opposite limit of $t_{e}/t_{a}>1$.
Fig.~\ref{fig:trigonalband2} shows the evolution of the band structure
as we increase $t_{e}/t_{a}$.
Notice that
the band gap at the $\Gamma$ point reduces
progressively as $t_{e}/t_{a}$ increases.
In particular, when $t_{e}/t_{a}$=$(t_{e}/t_{a})_{c}$=2.3,
a band inversion occurs at the $\Gamma$ point and the system
becomes metallic. The double degeneracy at the $\Gamma$ point
is protected by the lattice point group symmetry.
Therefore the metallic phase is stable as long as
symmetry breaking fields reducing the lattice symmetry are not introduced.
It is interesting that a similar metallic state
is predicted by a recent first principle calculation
on Y$_{2}$Ir$_{2}$O$_{7}$.~\cite{heungsik}
Although we have used a simplified tight-binding approach,
the overall band structure and degeneracies at the $\Gamma$
point for the metallic phase are consistent with the prediction of the LDA calculation.

%%%%%%%%%%%%%%%%%%%%%%%%%%%%%%%%%%%%%
\begin{figure}[t]
\centering
\includegraphics[width=8.5 cm]{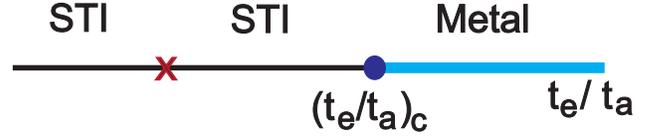}
\caption{
Phase diagram as a function of the strength of the local trigonal crystal field
represented by $t_{e}/t_{a}$.
Note that a metallic phase occurs when $t_{e}/t_{a}$ becomes larger than
the critical value of $(t_{e}/t_{a})_{c}>1$.
The red cross in the middle of two strong topological
insulators (STI) indicates the point where
accidental band touching occurs.
}
\label{fig:phasediagram}
\end{figure}
%%%%%%%%%%%%%%%%%%%%%%%%%%%%%%%%%%%%%%%%%%%

The effect of the trigonal crystal fields on the electronic
structure is summarized in the phase diagram shown in Fig.~\ref{fig:phasediagram}.
The strong topological insulator (STI) is stable against the trigonal crystal field
effect for $t_{e}/t_{a}$ smaller than the critical value, $(t_{e}/t_{a})_{c}$.
On the other hand, if $t_{e}/t_{a}$ is larger than the critical value,
a metallic phase occurs.
The metallic phase is stable as long as the point group symmetry of
the lattice is protected.

\section{\label{sec:LatticeDistortion} Lattice distortion and metal-insulator transition}
In this section we study the fate of the metallic phase predicted above
against external perturbations.
Some pyrochlore iridates A$_{2}$Ir$_{2}$O$_{7}$ with A= Nd, Sm, and Eu,
show metal insulator transitions as the temperature decreases.\cite{gingras,MIT}
However, the nature of the insulating ground state is under controversy.
According to a recent Raman scattering measurement on these iridium compounds,\cite{Raman}
metal-insulator transitions accompany
structural distortions for Sm$_{2}$Ir$_{2}$O$_{7}$ and Eu$_{2}$Ir$_{2}$O$_{7}$.
In addition, a recent theoretical study on a toy model for the topological insulators
on the pyrochlore lattice,\cite{franz} has shown that the lattice distortions
along the [111] direction may be important for the realization of the topological insulator.
Here we investigate the effect of lattice distortions
on the fate of the metallic state.
In particular, to make predictions relevant to future high pressure experiments,
we focus on uniform lattice modulations via $\textbf{q}=0$ phonon mode softening,
which lead to macroscopic volume change of the system.

%%%%%%%%%%%%%%%%%%%%%%%%%%%%%%%%%%%%%
%\begin{figure}[t]
%\centering
%\includegraphics[width=5.5 cm]{tetrahedronCoord.eps}
%\caption{
%A tetrahedral unit cell with 4 Ir atoms.
%Each Ir atom is represented by a black dot.
%Six bonds between neighboring Ir atoms are described by red solid lines.
%}
%\label{fig:tetrahedronCoord}
%\end{figure}
%%%%%%%%%%%%%%%%%%%%%%%%%%%%%%%%%%%%%%%%%%%

Since we have 4 Ir ions within a tetrahedral unit cell,
displacements of 4 Ir atoms from their equilibrium
positions, $\delta \textbf{r}_{i}=(\delta x_{i},\delta y_{i},\delta z_{i})$,
give rise to 12 independent degrees of freedom.
Among these 12 different modes, we neglect 6 modes describing global
translations and rotations of the unit cell,
because these do not lead to distortions of the tetrahedron.
The remaining 6 modes are classified as $A_{1}$ singlet $Q^{A}$,
$E$ doublet $\textbf{Q}^{E}=(Q^{E}_{1}, Q^{E}_{2})$,
and $T_{2}$ triplet $\textbf{Q}^{T}=(Q^{T}_{1},Q^{T}_{2},Q^{T}_{3})$,
in terms of irreducible representations for
the $T_{d}$ point group of the tetrahedron.~\cite{Tinkham, oleg1, oleg2}
The expressions for normal coordinates $Q$ corresponding to each irreducible representation
are given by
\begin{align}\label{eq:normalmodes}
Q^{A} & =\frac{1}{\sqrt{6}}(\delta r_{14}+\delta r_{23}+\delta r_{24}+\delta r_{13}+\delta r_{34}+\delta r_{12})
\nonumber\\
Q^{E}_{1} & =\frac{1}{\sqrt{12}}(\delta r_{14}+\delta r_{23}+\delta r_{24}+\delta r_{13}-2\delta r_{34}-2 \delta r_{12})
\nonumber\\
Q^{E}_{2} & =\frac{1}{2}(-\delta r_{14}-\delta r_{23}+\delta r_{24}+\delta r_{13})
\nonumber\\
Q^{T}_{1} & =\frac{1}{\sqrt{2}}(\delta r_{14}-\delta r_{23})
\nonumber\\
Q^{T}_{2} & =\frac{1}{\sqrt{2}}(\delta r_{24}-\delta r_{13})
\nonumber\\
Q^{T}_{3} & =\frac{1}{\sqrt{2}}(\delta r_{34}-\delta r_{12})
\end{align}

In the above $\delta r_{ij}$ indicates the change of the distance
between the $i$th and $j$th Ir atoms.
Since the $Q^{A}$ mode describes a uniform elongation or contraction
of all bond lengths, it does not change the symmetry of the unit cell.
So we neglect the $Q_{A}$ mode in the following discussion.
On the other hand, the two components of the doublet
$\textbf{Q}^{E}=(Q^{E}_{1}, Q^{E}_{2})$
describe tetragonal and orthorhombic distortions, respectively.
Finally, each component of the triplet mode, $Q^{T}_{i}$ depicts
elongation and contraction of a pair of orthogonal bonds
which are lying on two parallel planes of a cube. (See Fig.~\ref{fig:pyrochlorestructure}(b).)
Equal amplitude superposition of three components of the triplet mode
leads to a trigonal distortion of a tetrahedron along [111]
or its equivalent directions.

Using the information about the phonon modes of a single tetrahedron,
we consider various $\textbf{q}=0$ phonon modes on the pyrochlore lattice system.
The pyrochlore lattice consists of two inequivalent tetrahedra sharing a corner
and these two types of tetrahedra are interchanged via inversion symmetry
with respect to a corner.
Therefore even when we are restricted to $\textbf{q}=0$ phonon modes,
there are lattice distortions with even and odd symmetries with respect to an inversion center.
However, in this work, we focus on phonon modes with even inversion parities.
This is because only the phonon modes with even parities
lead to macroscopic distortion of the lattice.
Therefore these modes can be softened via coupling to applied external pressure,
which can be performed in future high pressure experiments.
%%%%%%%%%%%%%%%%%%%%%%%%%%%%%%%%%%%%%
\begin{figure}[t]
\centering
\includegraphics[width=6.5 cm]{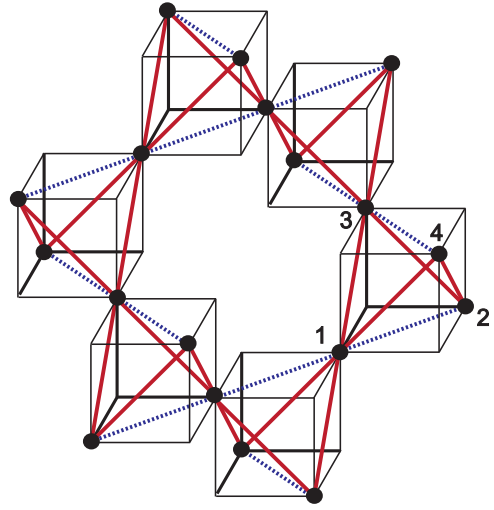}
\caption{
Tetragonal lattice distortion induced by a $\textbf{q}=0$ phonon mode
with $E_{g}$ symmetry. Red solid and blue broken lines
describe two inequivalent bond lengths under the lattice distortion.
}
\label{fig:tetragonalphonon}
\end{figure}
%%%%%%%%%%%%%%%%%%%%%%%%%%%%%%%%%%%%%%%%%%%
%%%%%%%%%%%%%%%%%%%%%%%%%%%%%%%%%%%%%
\begin{figure}[t]
\centering
\includegraphics[width=6.5 cm]{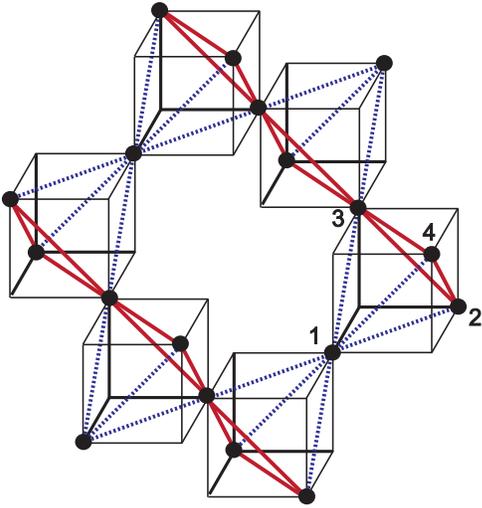}
\caption{
Trigonal lattice distortion induced by a $\textbf{q}=0$ phonon mode
with $T_{2g}$ symmetry. Red solid and blue broken lines
describe two inequivalent bond lengths under the lattice distortion.
}
\label{fig:trigonalphonon}
\end{figure}
%%%%%%%%%%%%%%%%%%%%%%%%%%%%%%%%%%%%%%%%%%%
In particular we focus on tetragonal and orthorhombic distortions
induced by $E_{g}$ phonon modes and trigonal distortions driven by
$T_{2g}$ phonon modes. Structures of distorted lattices and modulations
in bond lengths
are described in Fig.~\ref{fig:tetragonalphonon} and ~\ref{fig:trigonalphonon}.

Modulation of interatomic distances between neighboring Ir atoms
results in the renormalization of the nearest neighbor hopping amplitudes.
To understand the effect of electron-phonon coupling
on the ground state properties of the system,
we consider a Hamiltonian given by,
\begin{align}\label{eq:Helph}
H_{\text{el-ph}}=&\sum_{i,n,\alpha}(\varepsilon_{\alpha}-\mu)d^{\dag}_{in\alpha}d_{in\alpha}
+\frac{1}{2}\sum_{i}\sum_{n,m}K_{n,m}\eta^{2}_{n,m}
\nonumber\\
&+\sum_{\langle ij \rangle}\sum_{n,n'}\sum_{\alpha,\alpha'}
d^{\dag}_{in\alpha}T_{n\alpha,n'\alpha'}(\{\eta_{n,n'}\})d_{jn'\alpha'} + h.c.,
\end{align}
where $\eta_{n,n'}$ indicates modulation of the hopping amplitude
between neighboring $n$th and $n'$th Ir atoms.
Specifically,  we assume that $T_{n\alpha,n'\alpha'}(\{\eta_{n,n'}\})$
=$T_{n\alpha,n'\alpha'}(\{\eta_{n,n'}=0\})(1-\eta_{n,n'})$.
This is equivalent to scaling of $t_{a}$ and $t_{e}$ to
$t_{a}(1-\eta_{n,n'})$ and $t_{e}(1-\eta_{n,n'})$ between neighboring sites $n$ and $n'$.
(See Eq.(\ref{eq:Ttilde}) and (\ref{eq:Tfinal}).)
Since $t_{a}\propto t_{e} \propto \alpha_{pd\pi}^2$,
this approximation captures the change in the overlap integral between
neighboring $d$ and $p$ orbitals
caused by electron-phonon coupling.
The elastic constant, $K_{n,n'}$, corresponding to the modulation $\eta_{n,n'}$,
is simply taken to be $K_{n,n'}$ = 1.7$t_{a}$.

%%%%%%%%%%%%%%%%%%%%%%%%%%%%%%%%%%%%%
\begin{figure}[t]
\centering
\includegraphics[width=6.5 cm]{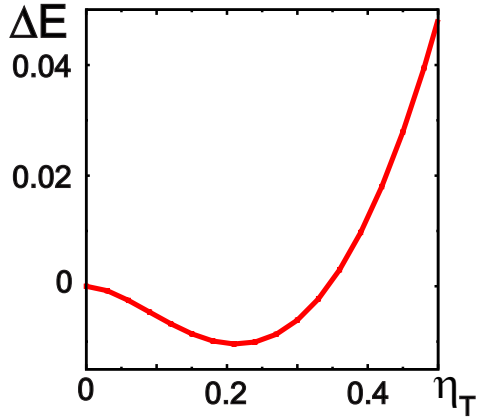}
\caption{
Ground state energy per unit cell for the pyrochlore iridate system coupled
with a trigonal phonon mode along [111] direction.
Here $\lambda_{SO}$=4.0, $t_{a}$=0.5, and $t_{e}/t_{a}$=2.5.
We plot the change of the ground state energy induced by the trigonal lattice distortion as a function
of the hopping amplitude modulation $\eta_{T}$.
}
\label{fig:phononenergy}
\end{figure}
%%%%%%%%%%%%%%%%%%%%%%%%%%%%%%%%%%%%%%%%%%%

\begin{figure}[t]
\centering
\includegraphics[width=8.5 cm]{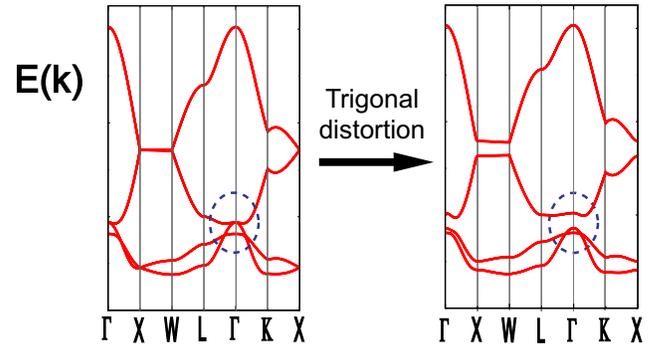}
\caption{
Band structure change induced by a trigonal lattice distortion,
which comes from a $\textbf{q}=0$ $T_{2g}$ phonon mode softening.
A band gap opens at the $\Gamma$ point.
The resultant insulating phase is a strong topological insulator.
}
\label{fig:trigonalphononband}
\end{figure}

In Fig.~\ref{fig:phononenergy} we plot the ground state energy
of the coupled electron-phonon system (Eq.(\ref{eq:Helph}))
as we increase the magnitude of hopping amplitude modulation $\eta_{T}$,
which corresponds to a trigonal lattice distortion.
Here $\eta_{T}>0$ means that the hopping amplitudes along the bonds connected with the site $1$
(the broken lines in Fig.~\ref{fig:trigonalphonon}) are reduced by $\eta_{T}$
while the hopping amplitudes along all other bonds are increased by the same amount.
In Fig.~\ref{fig:trigonalphononband}, we plot the change of the electron
band dispersion induced by a trigonal lattice distortion.
The trigonal lattice distortion results in opening a full gap at the Fermi energy
leading to an insulating phase.
The competition between the electronic energy gain from gap opening
and elastic energy cost compromises at the equilibrium bond distance.
Straightforward calculation of $Z_{2}$ topological invariants shows that
the resultant insulating ground state is a strong topological insulator
with $Z_{2}$ invariants ($\nu;\nu_{1}\nu_{2}\nu_{3}$)=(1;000).
Notice that the $Z_{2}$ invariants of the new insulating phase
are the same as those of the original topological insulating phase,
which exists when there is no local trigonal crystal field splitting effect.

%%%%%%%%%%%%%%%%%%%%%%%%%%%%%%%%%%%%%
%%%%%%%%%%%%%%%%%%%%%%%%%%%%%%%%%%%%%%%%%%%

It is interesting that two insulating phases separated by a metallic
phase in between, share the same topological properties.
To understand the reason for the identical topological properties of two insulating
phases, we have checked the inversion parities of all bands at the $\Gamma$ point
displayed in Table~\ref{table:bandparity}.
According to Fu and Kane,\cite{Fu-Kane} for a system with time-reversal and inversion symmetries,
the $Z_{2}$ topological invariants are given by the product of inversion parities
of all occupied bands at time reversal invariant momenta.
As one can see in Fig.~\ref{fig:trigonalband2},
the metallic phase induced by local trigonal crystal field effect has
a band crossing only at the $\Gamma$ point.
Since every band has even inversion parity at the $\Gamma$ point,
the band crossing does not change the parities of occupied bands.
Therefore two insulating phases with an intervening metallic phase
share the same $Z_{2}$ topological indices in this case.

\begin{table}
\begin{tabular}{@{}|c|c|}
\hline \hline
Lattice distortion & Ground state\\
\hline \hline
Tetragonal ($\eta_{E1}>0$) & Topological Insulator\\
Tetragonal ($\eta_{E1}<0$) & Metal  \\
\hline
Orthorhombic ($\eta_{E2}>0$) & Metal  \\
Orthorhombic ($\eta_{E2}<0$) & Metal\\
\hline
Trigonal ($\eta_{T}>0$) & Topological insulator\\
Trigonal ($\eta_{T}<0$) & Metal  \\
\hline \hline
\end{tabular}
\caption{Lattice distortions and the resulting ground states.}
\label{table:phonon}
\end{table}

In addition to trigonal lattice distortions,
we have also investigated the effect of tetragonal and orthorhombic distortions,
which result from softening $\textbf{q}=0$ $E_{g}$ phonon modes.
It turns out that a tetragonal distortion with
$\eta_{E1}>0$ leads to an insulating ground state.
Fig.~\ref{fig:tetragonalphonon} describes
a bond modulation pattern for
a tetragonal distortion with $\eta_{E1}>0$.
Here electron hopping amplitudes are increased for dotted bonds
while the hopping amplitudes for solid bonds
are reduced. Relative magnitude of bond length modulations
is consistent with the $Q^{E}_{1}$ mode, the first component of the doublet
$E$ phonon mode in Eq.(\ref{eq:normalmodes}).
However, the magnitude of the band gap generated by
a tetragonal distortion with
$\eta_{E1}>0$ is much smaller than that from a trigonal lattice distortion
when the magnitudes of hopping amplitude modulations are the same in the two cases.
In addition, an orthorhombic distortion does not open a full band gap.
Therefore we propose that application of external pressure along the [111]
or its equivalent directions is the most efficient way of generating
strong topological insulators.

\section{\label{sec:conclusion} Conclusion}

We investigate the possible existence of the strong topological
insulators in pyrochlore iridates A$_{2}$Ir$_{2}$O$_{7}$.
The effective Hamiltonian for the pyrochlore lattice of Ir ions
is derived by taking into account the strong spin-orbit
coupling and trigonal crystal field effect.
It turns out that the strong topological insulator
found for the ideal cubic environment of Ir ions
may turn into a metallic state under trigonal distortion of oxygen octahedra.
It has been known that, in cubic pyrochlore oxides, A$_{2}$B$_{2}$O$_{7}$,
where both A and B site ions reside on two distinct interpenetrating pyrochlore networks,
the trigonal crystal field splitting exists inherently~\cite{Structure_Review}
with an energy scale comparable to the spin-orbit coupling.~\cite{heungsik}
This may also be consistent with the recent LDA calculation of Y$_{2}$Ir$_{2}$O$_{7}$
where it was found that the non-magnetic ground state would be a metal.~\cite{wan,heungsik}
Given that various pyrochlore iridates with different A-site ions
possess substantial but different amount of trigonal distortion,
the presence of trigonal crystal field effect may be one of the important factors
that determine the nature of low temperature ground states and finite temperature
metal-insulator transitions.

On the other hand, we found that the electron-lattice coupling also plays
an important role. It is shown that certain $\textbf{q}$=0 normal modes
lead to the re-emergence of a strong topological insulator
when these modes are softened and the system undergoes a structural
deformation.
Recent experimental observation of
the pressure-induced metal-insulator transition in Ba$_{1-x}$R$_{x}$IrO$_{3}$ (R=Gd, Eu)
suggests that electron-lattice coupling strongly affects
the ground state properties in iridium oxide compounds.
Due to the sensitive response of the electronic structure near the Fermi level against the variation of
Ir-O-Ir bond angles, the application of moderate hydrostatic pressure around 12 kbar
destabilizes the metallic ground state leading to metal-insulator transition.~\cite{BaIrO}
%Our study may also explain the recent experimental discovery
%of the structural transition in certain prochlore iridate systems, accompanying
%the finite temperature metal-insulator
%transition.
While the identification of the true ground state in the pyrochlore iridates
would require better understanding of the electron correlation effect, it is an
intriguing possibility that
the strong topological insulators may arise via a finite temperature
metal-insulator transition with structural changes.
Our study also suggests
that the application of the external pressure along [111] or its equivalent directions
may lead to a strong topological insulator by taking the advantage of the
electron-lattice coupling or perhaps even to a topological Mott insulator in
stronger correlation regime.

In the current work, we did not study the electron-electron interaction
effect. It is possible that sufficiently strong electron interaction would turn the metallic
state induced by the trigonal crystal field effect to a magnetically
ordered insulator or a more subtle form of Mott insulator.
The competition between magnetically ordered Mott insulators, topological
band insulators, and topological Mott insulators in the presence of
electron interaction would be an excellent topic of future studies.
The understanding of the delicate interplay between the electron interaction,
spin-orbit coupling, local crystal field effect, and electron-lattice coupling would
be essential for the determination of the ultimate ground states in these systems.
All of these possibilities await for further experimental verifications and findings.

\acknowledgments

We thank Heungsik Kim and Jaejun Yu for insightful discussions.
This work was supported by the NSERC of Canada, the Canada Research Chair
Program and the Canadian Institute for Advanced Research.

%%%%%%%%%%%%%%%%%%%%%%%%%%%%%%%%%%%%%%%%%%%%%%%%%%%%%%%%%%%%%%%%%%%%%

\appendix

\section{Explicit matrix expressions for the effective hopping Hamiltonian }

In this appendix, we present the expressions for various matrices
used to derive the effective hopping Hamiltonian in Sec.~\ref{sec:H_hopping}.
We follow the same convention taken by Pesin and Balents.\cite{Pesin-Balents}
The matrix $R^{(i)}$ rotating the global cubic axis to the local
octahedral axis for the $i$th Ir atom is given by

\begin{displaymath}
R^{(1)}=
\left( \begin{array}{ccc}
\frac{2}{3} & -\frac{1}{3} & -\frac{2}{3}  \\
-\frac{1}{3} & \frac{2}{3} & -\frac{2}{3}  \\
\frac{2}{3} & \frac{2}{3} & \frac{1}{3}
\end{array} \right),
R^{(2)}=
\left( \begin{array}{ccc}
\frac{2}{3} & \frac{2}{3} & \frac{1}{3}  \\
-\frac{2}{3} & \frac{1}{3} & \frac{2}{3}  \\
\frac{1}{3} & -\frac{2}{3} & \frac{2}{3}
\end{array} \right),
\end{displaymath}

\begin{displaymath}
R^{(3)}=
\left( \begin{array}{ccc}
\frac{1}{3} & -\frac{2}{3} & \frac{2}{3}  \\
\frac{2}{3} & \frac{2}{3} & \frac{1}{3}  \\
-\frac{2}{3} & \frac{1}{3} & \frac{2}{3}
\end{array} \right),
R^{(4)}=
\left( \begin{array}{ccc}
\frac{1}{3} & -\frac{2}{3} & \frac{2}{3}  \\
-\frac{2}{3} & -\frac{2}{3} & -\frac{1}{3}  \\
\frac{2}{3} & -\frac{1}{3} & -\frac{2}{3}
\end{array} \right).
\end{displaymath}

The matrix $D^{(i)}$ corresponding to a spinor representation of $R^{(i)}$
is given by

\begin{displaymath}
D^{(1)}=
\left( \begin{array}{cc}
\sqrt{\frac{2}{3}} & \frac{1-i}{\sqrt{6}} \\
-\frac{1+i}{\sqrt{6}} & \sqrt{\frac{2}{3}}
\end{array} \right),
D^{(2)}=
\left( \begin{array}{cc}
-\frac{2+i}{\sqrt{6}} & -\frac{i}{\sqrt{6}} \\
-\frac{i}{\sqrt{6}} & -\frac{2-i}{\sqrt{6}}
\end{array} \right),
\end{displaymath}

\begin{displaymath}
D^{(3)}=
\left( \begin{array}{cc}
\frac{2-i}{\sqrt{6}} & -\frac{1}{\sqrt{6}} \\
\frac{1}{\sqrt{6}} & \frac{2+i}{\sqrt{6}}
\end{array} \right),
D^{(4)}=
\left( \begin{array}{cc}
-\frac{i}{\sqrt{6}} & \frac{1-2i}{\sqrt{6}} \\
-\frac{1+2i}{\sqrt{6}} & \frac{i}{\sqrt{6}}
\end{array} \right).
\end{displaymath}

Finally, we show the matrix $A$ which changes the trigonal basis
to local spin-orbit eigenstates as follows.
\begin{displaymath}
A=
\left( \begin{array}{cccccc}
0 & 0 & 0 & 0 & \frac{n_{1}\omega}{3} & 0 \\
0 & 0 & 0 & 0 & 0 & \frac{-n_{1}\omega}{3} \\
\frac{-n_{2}}{6} & 0 &  \frac{-i n_{2}\omega}{6} & 0 & 0 & 0 \\
0 & \frac{-n_{2}}{3\sqrt{2}} & 0 & \frac{-i n_{2}\omega}{3\sqrt{2}} & 0 & 0 \\
\frac{n_{3}}{3\sqrt{2}} & 0 & \frac{-i n_{3}}{3\sqrt{2}} & 0 & 0 & 0 \\
0 & \frac{n_{3}}{\sqrt{6}} & 0 & \frac{-i n_{3}}{\sqrt{6}} & 0 & 0
\end{array} \right),
\end{displaymath}
where $\omega=e^{i\frac{2\pi}{3}}$ and
$[n_{1}n_{2}n_{3}]$ indicates the local $C_{3}$ symmetry axis
of the corresponding distorted oxygen octahedron.

%%%%%%%%%%%%%%%%%%%%%%%%%%%%%%%%%%%%%%%%%%%%%%%%%%%%%%%%%%%%%%%%%%%%%%%%%%%%%%%%%%%%%%%%

%%%%%%%%%%%%%%%%%%%%%%%%%%%%%%%%%%%%%%%%%%%%%%%%%%%%%%%%%%%%%%%%%%%%%%%%%%%%%%%%%%%%%%%%%%%

\end{document}